**(Note: Originally 2007; Section 3.3.5 revised 1/2011)**

---------------------------------------------------------------------------------------------

# Improving TBI Protection Measures and Standards for Combat Helmets

**Eric G. Blackman**
Department of Physics and Astronomy, University of Rochester, Rochester, NY, 14627

(Report I from TBI DSSG 2006-7 Discussion Team composed of Eric G. Blackman (a), Melina E. Hale (b) Sarah H. Lisanby (c)
a. Department of Physics and Astronomy, University of Rochester, Rochester, NY, 14627
b. Department of Neuroscience, Columbia University, New York, NY, 10032
c. Department of Organismal Biology and Anatomy, University of Chicago, Chicago, IL, 60637)

**1. ABSTRACT**

How well do current combat helmets protect against Traumatic Brain Injury? To answer this question it is necessary to evaluate both the *measures* and *standards* of protection. We define *measure* as a physical test that the helmet must be subject to and the *standard* as the quantitative threshold of performance in this test that the helmet must satisfy to be acceptable. We find that both the measures are inadequate and the standards are too low. Current combat helmets simply do not protect against TBI effectively as they have not been designed to do so. We explain this inadequacy of combat helmet protection against TBI, and how the situation can be dramatically improved both in the short and long term.

**2. Introduction**

As emphasized in Part I, and Part II, Traumatic Brain Injury (TBI) refers to a clinical syndrome of impairments in brain function resulting from physical brain trauma. TBI is not necessarily accompanied by skull fracture or visible external head injury and for these cases the condition is referred to as closed TBI. This is the primary focus herein.

Approximately 1.5-2 million civilian TBI cases have been diagnosed each year since the early 1990s, leading to about 200 hospitalizations per year per 100,000 people and about 56,000 deaths per year. This also corresponds to a death rate of ~20 per 100,000 people (McArthur et al. 2004). About 50% of TBI cases are caused by automobile accidents and approximately 20% are sports related injuries (Bohnen et al. 1992). The majority of TBI cases is non-fatal and classified as mild traumatic brain injury (MTBI) or concussion. The rate has halved from 1980 (Kraus and McArthur 1996) but seems to have remained steady since the late 1990s. An increase in MTBI from recreational sports has also been recognized.

The TBI problem is acute for military personnel as exemplified even by the medical cost alone: Through 2007, there have been approximately 1.4 million soldiers serving in Iraq and Afghanistan. About 3.6% return with injuries, and approximately 60% of injured soldiers have some form of MTBI or TBI, and about 20% of these with serious

TBI (about ~10000 soldiers through early 2007.) The cost of treatment for per warfighter with serious TBI is estimated between $600,000 and $5,000,000 per such patients over their lifetimes (Wallsten & Kostec 2005; Bilmes & Stiglitz 2008, Stiglitz and Bilmes 2008). If we assume that the 10,000 soldiers with serious TBI injured between 2003-2007 all live for 25 years, and estimate an intermediate average lifetime care cost of 2.7 million for each, then the total cost from these soldiers alone would be about 27 billion and with an annual cost of 1.1 billion. This does not include future injured, soldiers, lost wages, vocational costs, adaptive housing expenses, and disability compensation beyond medical treatment. This estimate is consistent with those of Bilmes & Stiglitz (2008) .

The situation may prove to be even worse as screening becomes increasingly more effective. According to Charles Dasey, spokesman for the U.S. Army Medical Research and Materiel Command (e.g. Emery 2007), Recent TBI specific screenings at Fort Carson revealed that about 18 percent of all troops who returned from Iraq in 2006 and 2007 from 13,400 soldiers screened had brain injuries. Of those, 13 percent were unfit to return to Iraq. Similarly, between 10% and 20% of troops returning to Camp Pendleton in California, Fort Bragg in North Carolina and Fort Hood in Texas incurred TBI. If more than 10% of all troops have some form of TBI, this would imply more than 140,000 troops from the start of the Iraq conflict through early 2007. If 10% of those have serious TBI, this would imply 14,000 troops, raising the cost estimate from $27 billion above to $38 billion just to treat the troops with TBI so far. For a 25 year life span of soldiers with TBI, this would already imply a $1.5 billion annual cost. The multifaceted human and financial costs of TBI highlight the need to better understand, detect, prevent, and treat TBI.

Here we focus specifically on the role of helmets and the principles that need to be incorporated in helmet design to protect against TBI. Combat soldiers incur TBI most commonly from exposure to blasts. However, blasts produce closed TBI from both overpressure and from head impact because a blast force that is strong enough to cause non-fatal overpressure injury will easily propel a soldier to the ground or into an obstacle. There is presently no TBI measure or standard for helmets to protect against overpressure, and only a crude TBI standard for impact injury. We will see that this measure is inadequate but that military helmets still fail to meet it.

To understand the ineffectiveness of present helmets, it is first necessary to understand the basic physics of how the brain can be injured. In section 3.1 we discuss the physics of brain injury from impacts, which has traditionally been the focus of helmet protection for blunt impact. In section 3.2 we discuss the origin and identify inadequacies of the current helmet measures used for combat helmets. In section 3.3, lessons and insight from motor vehicle, military and football helmet studies are discussed. We find that the football helmet research community leads the way with respect to the most dramatic improvements in helmet design and real time injury diagnostics. In section 3.4 we describe recent progress toward understanding the distinct effects of blast overpressure vs. impact injury via numerical simulations and where important future directions for this research. Section 3.5 presents conclusions and specific recommendations.

## 3.1 Brain Injury from Impact

The brain is coupled to the skull viscously via the surrounding cerebral spinal fluid (CSF). During normal activity, the viscous force damps the brain motion, keeping it from impacting the skull. However, for a sufficiently large external force associated with a rapid deceleration, the viscous coupling between the brain and skull is insufficient, and the brain will crash against the skull wall. More specifically, consider a helmeted head with a large velocity that impacts an immobile surface, as might occur if a solider is knocked over and hits his/her head on the ground. The helmet will protect the skull from fracture but the rapid deceleration from the maximum speed to zero upon impact exerts a strong force of deceleration. The viscous coupling between head and brain cannot stop the brain when the head impacts; the brain crashes into the interior skull wall and deforms. Different manifestations of TBI result from this brain deformation and the associated tissue stress. Because there is a much lower threshold force to injure the brain compared to the skull, skull fracture need not accompany physical brain damage and the resulting head injury is termed "closed."

The TBI-inducing brain deformation just described requires energy. This energy ultimately comes from gravitational and/or chemical energy of an explosion, which gets converted into bulk kinetic energy of the soldier. Once the moving head (and body) impact a stationary obstacle, the kinetic energy is converted into deformation energy of the head form. Some of this deformation energy is absorbed by the helmet and skull, but brain damage results from the residual energy absorbed by the brain itself. Ideally, the more energy that can be absorbed by the helmet, the less available energy there is for injury.

Energy absorption is not, however, the only role that the helmets must play to protect against TBI. The localized forces on brain tissue must also be minimized, even if the total energy left for the brain is small. This implies that the force to the head must be minimized. A large amount of energy applied to the brain over a long time with a weak force may do no damage whereas a small amount of energy imparted abruptly with a very strong force may tear brain tissue. Both the total energy and the rate of energy deposition to the brain (or equivalently, the total energy and the force) must ultimately be incorporated in determining the threshold for TBI.

To understand the role of force vs. energy, consider an analogy to everyday experience: gently braking a car to a stop from highway speeds does no damage to a passenger while removing all of the kinetic energy of bulk motion. Now consider the car and passenger to be moving at the same highway speed, but suddenly crash into a wall. Passengers not wearing seatbelts will get injured as they get thrown and crash inside the vehicle. This illustrates the importance of slowing the rate of energy change, which in turn implies slowing the rate of velocity change i.e. making the magnitude of the deceleration as small as possible. Because force equals mass times acceleration (or deceleration) a small deceleration magnitude reduces the external force on the head form, which in turn, reduces the internal forces on the internal brain tissue. In the automobile, the passenger is analogous to the "brain" and the internal passenger compartment is analogous to the inside of the skull. The bumper and fenders are analogous to the helmet cushioning and shell

To protect against impact induced TBI, a well-designed helmet must therefore both absorb energy from the impact, leaving less energy for the skull and brain, and also cushion the impact to minimize the magnitude of deceleration. Although a hard shell protects against ballistic impact and localized skull fracture, without appropriate cushioning somewhere between the impact site and the skull, the helmet is ineffective against protecting closed head brain injury. Well-designed cushioning increases the time scale over which the head decelerates from its terminal speed to zero speed upon impact. It may also be possible to design the shell of the helmet to be relatively "soft" while still protecting against shrapnel and projectile impact.

An immediate goal of helmet design should be to identify and require better impact protection cushioning and an important longer term goal should be to develop correlations between external and internal measures of TBI, where external refers to forces on the head form and internal refers to the resulting localized forces and stresses on the brain tissue. This correlation will come from combining the analytic physical principles summarized above with comprehensive high resolution numerical simulations that can follow the dynamics of external forces on a head form and the internal brain stresses that result. We will return to discuss such simulations later.

## 3.2 Origin and Inadequacy of Current TBI Helmet protection measures

Helmet protection against skull fracture from the blunt impact is a less demanding requirement than protection against TBI. Some helmets do incorporate a blunt impact standard for skull fracture but this will insufficiently protect against TBI. As discussed below, the only measure currently used to specifically address TBI threshold is based on pre 1980 data from monkey concussions scaled to human heads and only the National Operating Committee on Standards in Athletic Equipment (NOCSAE) employs a blunt impact standard for concussion protection by football helmets. Presently, neither motor vehicle helmets nor military helmets require a TBI standard. We describe the origin and limitations of the measures, and the ineffectiveness of the standards below.

Helmet safety measures currently used for protection against blunt impact trauma are all based on mitigating the magnitude of the deceleration of a helmeted head form drop tests onto hard surfaces in laboratory tests. As an object falls to the ground, its speed reaches a maximum just before impact. Once it hits the impact surface, the magnitude of deceleration rises from zero to a maximum, and then falls back to zero as the object comes to rest. The force imparted to the head form at any one time during the impact is proportional to the instantaneous deceleration. As discussed above, decreasing the peak value of the deceleration and decreasing the time over which this acceleration lasts, both reduce the stress on the head and brain. Figure 3.1 shows the characteristic evolution of acceleration as a function of time and highlights the conceptual distinction between peak acceleration and the duration of acceleration above some threshold.

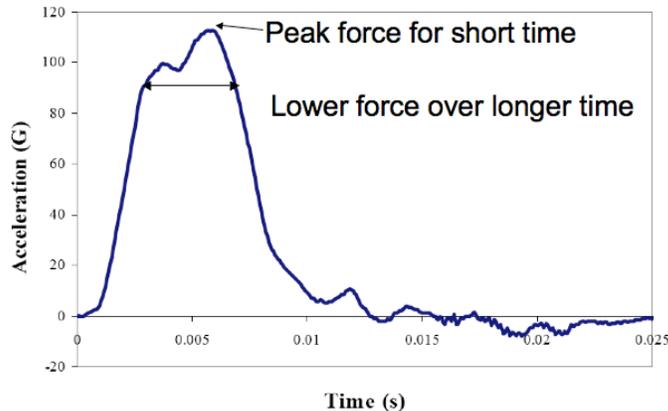

Figure 3.1: A typical time evolution curve for an acceleration (=magnitude of deceleration) of a helmeted headform subjected to a drop. The curve is meant for illustrative purposes but is roughly consistent with those from a 6kg head form with a currently used PASGT padded helmet system dropped from 3 feet. (e.g. Fig M-18 of Trimble et al. 05)

A measure that does play a role in TBI is the peak head form acceleration. One design goal is to reduce the peak acceleration during impact that reduces the peak force for a head form of fixed mass. However, as emphasized above, brain injury can also occur even if a sub-peak acceleration acts over a longer time period that that of the peak ( Fig 3.1). Therefore, some combination of acceleration and its duration must be constructed. In the helmet industry, the need to incorporate a combination of acceleration and duration arose not from purely theoretical considerations, but from empirical data produced by the auto crash testing that was fit with the Wayne State Tolerance Curve (WTSC) (Pattrick et al. 1963, Snyder 1970). The WSTC is a purely empirical graphical curve (see Fig 3.2) indicating the maximum value of deceleration at a given duration that can supposedly be tolerated without severe brain injury, originally considered as equivalent to the standard for skull fracture. The data from which the original curve derived came from (1) drop tests of 4 embalmed cadaver heads on plates, with measurements of linear acceleration, intracranial pressure and skull damage (2) air blasts to exposed cadaver brains and (3) hammer blows to animals. The curve highlights that that the threshold for head injury depends on the combination of the two variables such that weak decelerations can be tolerated for longer durations than large decelerations. Formulae based on this curve have become the dominant standard in current use for establishing a blunt impact protection threshold for head and brain injury.

The qualitative conclusions that lower accelerations can be tolerated for longer periods and that asymptotically, the threshold acceleration is independent of duration, are likely correct. However the WSHTC data are fundamentally flawed: The small number of data points, the dated instrumentation methods, the imprecision in definitions of quantities actually measured, the combined use of embalmed cadavers and animals, and the ambiguities between combining and interpreting date for open head injury (fracture) vs. closed head injury (non-fracture) injury into a human head injury/concussion curve have long been questioned (e.g. Versace 1971; Goldsmith

1981; Cory et al. 2001).

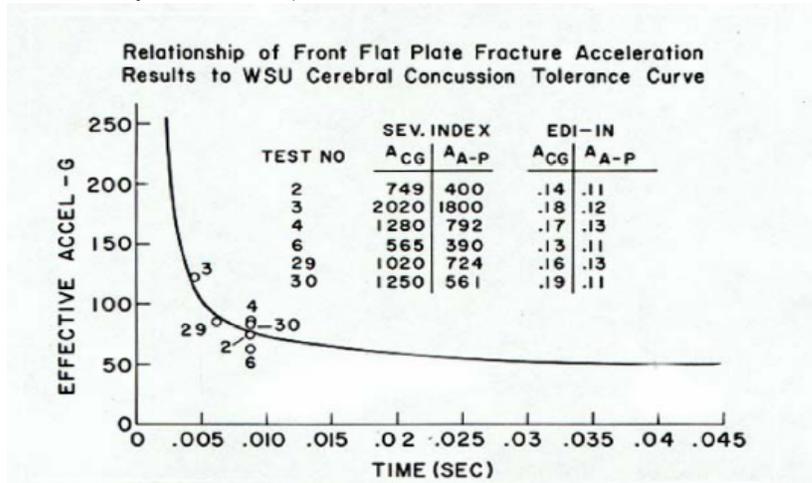

Figure 3.2: Wayne State Head Tolerance Curve (Patrcick 1963; Snyder 1970). Data are fit by the curve, which is extended well beyond the regime of the actual data.

Some improvement to the WSHTC curve was made by more clearly distinguishing skull fracture from concussions via Ono et al. (1980) as seen in Fig. 3.3.

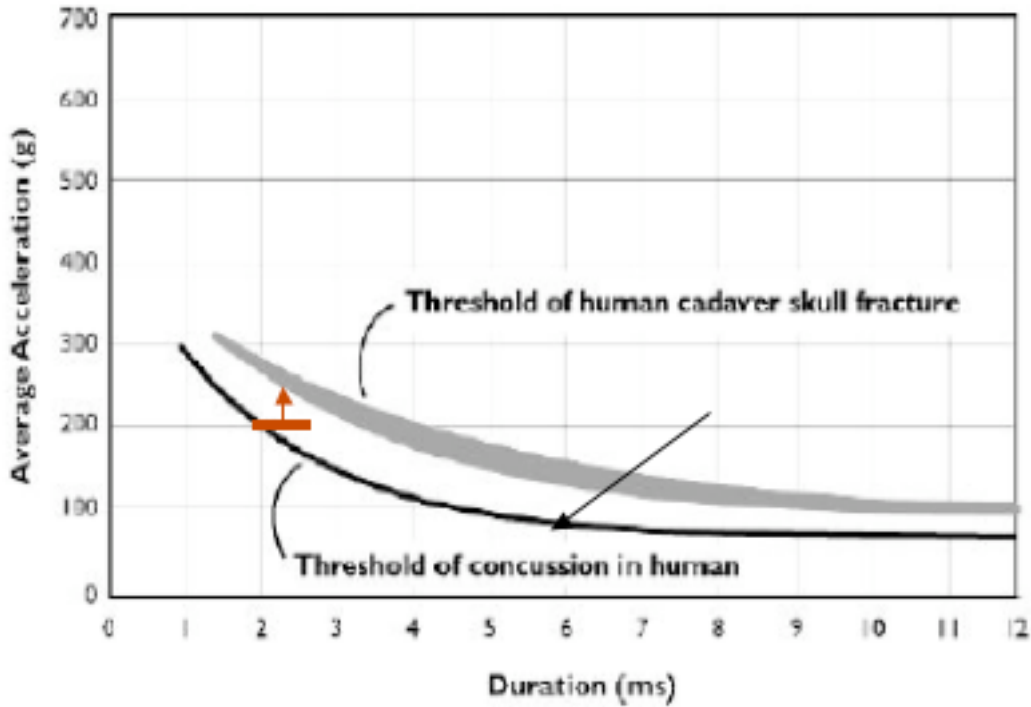

Figure 3.3: Japan Head Tolerance Curve (JHTC) proposed for concussions as distinguished from skull fracture threshold (Ono et al. 1980). Concussion threshold is based on head acceleration from head impact on live monkeys, then mass and surface area scaled to human heads via dimensional analysis, The actual data from which the curve is fit, span the range between 2-10ms. The horizontal line segment and arrow indicate roughly where current combat helmets would lie on this curve, when we use the data in Trimble et al. (2005) and McEntire et al. (2005).

Ono et al. (1980) performed experiments both with live monkeys and human cadavers and compared the threshold of skull fracture vs. concussion. The concussion severity was inferred in the post-impact monkeys via their loss of coronal reflex, duration of apnea, and blood pressure changes. The monkey data were then scaled to human heads using mass and surface area dimensional analysis. The data for fracture were then tested and compared with cadavers to validate the scaling relations used. The supposed resulting tolerance curve for human concussions is shown in Fig. 3.3. This threshold for concussion, in principle, would be an indication of mild TBI, and thus valuable. We will explain why this threshold is too lax below. However, it is alarming that even if the JHTC curve were the correct threshold, data from current military PAGST and ACH helmets (McEntire et al. 2005; Trimble et al. 2005) fitted with currently used padding inserts fall mostly above the horizontal line segment as indicated in Fig 3.3.

Ironically, the WSHTC and JHTC papers, on which so much subsequent head injury standardization has been based, did not appear in peer-reviewed journals but rather in automobile industry conference proceedings. That being said, even if the aforementioned problems with the data collection and analysis for the WSHTC could have been alleviated via peer review, an even more fundamental problem is that these data focus only on acceleration vs. time. Acceleration only uniquely determines the relative force for different impacts when the mass is kept fixed. In lab tests, the relevant masses can be kept fixed both by using standardized head forms and by using fixed body orientations if animals or dummies are used. But for realistic impacts, particularly in combat, the angle of impact that the body makes as the head impacts a hard surface can range from 0 to 90 degrees, where 0 represents a body parallel to the ground and 90 degrees is a perpendicular impact. For any angle of impact greater than 0, the effective mass of the head will be larger than just that of the head form. A rough calculation suggests that for obliquities of only 45 degrees the effective head mass can more than double for a typical 6ft male soldier of 200lbs which in turn, more than doubles the force. Brain injury is ultimately determined by the stress on brain tissue, and a larger effective mass implies a larger external force and more stress delivered to the brain. Having emphasized the importance of mass for realistic measures, we now return to the origin of the WSHTC fit formulae in more detail—for which the mass plays no role.

The Gadd (1961) severity index (SI) was the first attempt to incorporate the WSHTC into a practical formula. The GSI formula is given by the time integral

$$SI = \int_{t_1}^{t_2} a^n dt \tag{3.1}$$

where *a* is the acceleration in units of gravity, and *n* is an empirical index taken to be 2.5 to match the WSHTC and *t* is the time in seconds. Typically, values of ~1000 are chosen as the required threshold above which serious brain injury is expected based on the WSHTC or JHTC.

But the SI suffers from the problem that impacts of very slow deceleration extended over a long period give the same SI as high deceleration impacts of short duration. Common experience dictates that the head injury likelihood of the two cases would be quite different; for very small accelerations, one would expect no injury at all. This failure led to the Head Injury Criterion (HIC) as an attempt to refine the SI to integrate only over the part of the acceleration vs. duration curve that matters most for the impact. The HIC is given by

$$HIC = \left[(t_2 - t_1)\left(\frac{1}{t_2 - t_1}\int_{t_1}^{t_2} adt\right)^{5/2}\right]_{max} \quad (3.2)$$

where $a$ is the acceleration again in units of gravity and $t_2-t_1$ represents a chosen time interval (measured in seconds), in the deceleration vs. time curve that bounds the peak deceleration. The HIC is the most commonly used helmet safety standard for blunt impact, and for acceleration profiles with sharp rises, and flat plateaus, the SI and HIC are comparable. The NOCSAE uses a value of SI=1200 as the current threshold for football helmets. Different agencies using the HIC use different maximum allowed threshold values and different parameters to determine this threshold: The national highway traffic and safety administration (NHTSA) uses $t_2-t_1$ =36 milliseconds while the International Safety Organization (ISO) uses $t_2-t_1$ =15 milliseconds. The HIC between 500 and 2500 is typically converted into a probability for fatality by head injury (Fig 3.4); e.g. using 15 milliseconds, a 15% chance of fatal head injury for an HIC of 1000 has be claimed (Prasad and Mertz 1985). The arbitrariness in the choice of time interval, the absence of a mass in either the HIC or SI, and the limitations of the WSHTC and JHTC on which the HIC is based substantially reduce its reliability.

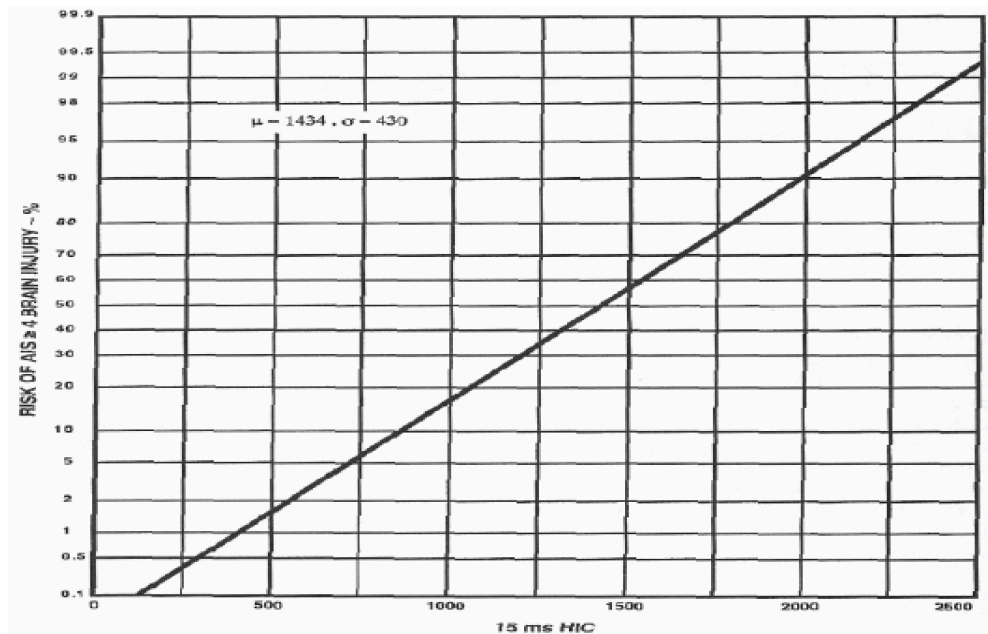

Figure 3.4: Typical injury risk curve associated with HIC. The vertical axis is the risk of injury and the horizontal axis is the HIC. The curve indicates the likelihood of concussion for a given HIC value (Prasad & Mertz 1985).

To further illustrate a serious limitation of the HIC and SI, we consider the revealing case of the acorn woodpecker. Gibson (2006) studied why woodpeckers do not get concussions and where the measured acceleration of woodpeckers' heads appear relative to the JHTC curve. The result is shown in Fig 3.5. The parallelogram on the plot indicates the measured accelerations (which are primarily linear acceleration) during impact. These data fall well above the JHTC. The likely explanation emerges by assuming that the stress (force per unit area) on the actual brain tissue required for injury is roughly the same for both woodpecker and humans and that the density of the brain tissue is also the same. Then the force per unit area delivered to the brain depends on the surface area and the mass. The ratio of the tolerable acceleration then scales roughly as the ratio of a woodpecker brain width to a human brain width. This leads to ~16 times larger tolerable acceleration for a given acceleration duration than for a human brain, explaining why woodpeckers do not get TBI as part of their daily routines. (As mentioned above, a similar procedure was used for scaling the original monkey data to construct the JHTC in the first place (Ono et al. 1980).

For present purposes, the woodpecker study highlights the role of the mass in the acceleration tolerance curve and the basic principle that the same stress on brain tissue can occur for very different accelerations when the mass of impacting object is varied. Most importantly, this even applies when comparing different impacts for similar humans under different circumstances. The relevant mass is not the brain mass, but the effective total mass lying along the direction of impact. This is a recurrent theme in our discussion. Ironically, the Gibson (2006) study makes an oversight in this regard. In that paper, only brain masses were used in scaling JHTC to establish the woodpecker tolerance curve of Fig 3.5. In actuality, the correct mass to scale would be the total effective head mass. Most likely, this revised scaling would lower the human JHTC and raise the woodpecker curve, exacerbating the difference between the two.

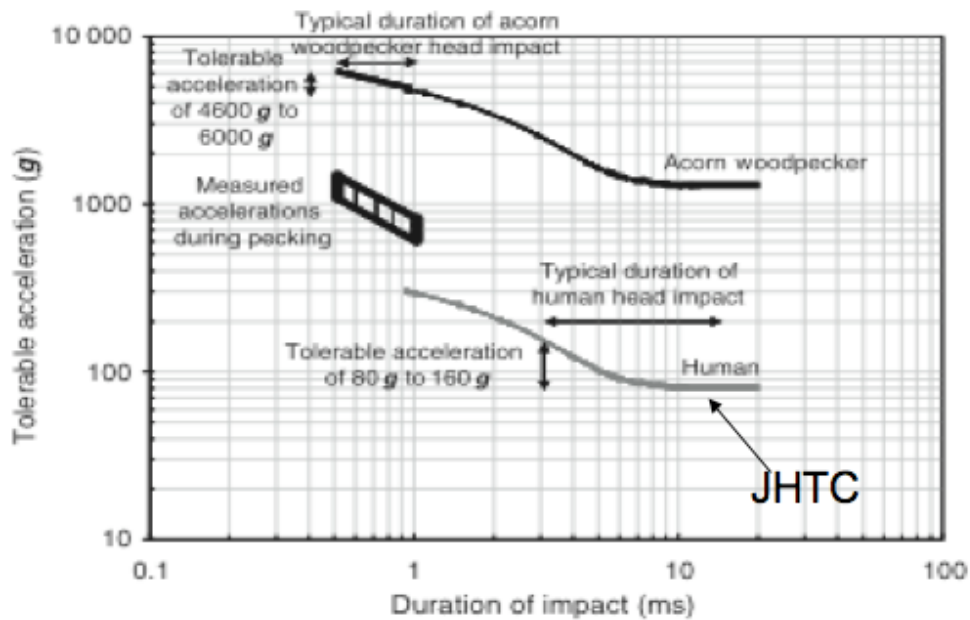

Figure 3.5: Comparison of JHTC and the analogous curve for the acorn woodpecker when the JHTC is scaled to a lower brain mass, assuming that woodpecker and human brain tissue have the same density and same stress thresholds for injury (Gibson 2006).

## 3.3. Lessons From Previous Helmet Studies and Standards

### 3.3.1 Motor vehicle helmets: Poor protection against TBI

The Fédédration Internationale de L'Automobile (FIA) Super Helmet Standard for auto racing does include an HIC threshold for blunt impact but the value is HIC = 3500. This is much too high to protect against closed head TBI even by the JHTC standard (Fig 3.3). The Snell SA2000 racing helmet standard for NASCAR has no HIC criterion. The auto industry has been more focused on skull fracture than closed TBI. In 2005, the Medical College of Wisconsin hosted a motor vehicle themed workshop on the "Criteria for Head Injury and Helmet standards." The discussion of the limited empirical value of the HIC as a helmet standard led the majority of attendees (by hand vote) to reject the HIC as a standard requirement for helmets (Fenner et al. 2005). Impact energy attenuation, crush protection, penetration protection, rotational acceleration protection, shell hardness, chin guard impact protection, all emerged as higher priorities. Indeed, imposing an HIC standard of 3500 is basically useless for closed TBI protection. However, imposing a more stringent value, ~150 (where football concussions are seen Fig 3.7) would require significant redesign of motor vehicle helmets.

Motorcycle helmets have somewhat more stringent standards for blunt impact but are still ineffective for closed TBI. Common standards are listed with the maximum allowed head acceleration (deceleration) in units of gravity and the duration of this acceleration where appropriate in parentheses (Thom et al. 1998): FMVSS 571.288 (150G, 200G, 400G for 4ms, 2ms and peak respectively) ANSI-Z90.1 (300G at peak acceleration) Snell 2000 (300G at peak acceleration). In general, the motorcycle helmets have equivalent HIC values that range from 1200 to 2400.

### 3.3.2. Combat Helmets: Combat helmets inadequate protection against TBI

Slobodnik (1980) recognized that the army air-crew helmet standard for acceleration was insufficient to protect against TBI. The specifications for this helmet allowed peak accelerations as high as 400G for a 6.3kg helmeted head form dropped from a height of 1.5 meters, corresponding to 95 Joules of impact energy. As a ``quick fix" Slobodnik recommended replacing 400G by 150G for a peak acceleration standard to protect against concussions, based on empirical correlations between helmet damage from military helicopter crashes and clinically recorded head injuries. This was eventually accommodated into the HGU-56/P aircrew helmet (McEntire et al. 2005).

Presently however, the only non-aviator helmet that has a blunt impact protection requirement is the Modular Integrated Communications Helmet (MICH) for special forces. Discordant with the implications of Slobodnik (1980) the threshold required for the MICH helmet is 150G at 10 feet per sec (fps) impact speed. This speed corresponds free-fall from a height of just 1.5 feet, not 1.5 meters.

Recognizing that the currently used Advanced Combat Helmet (ACH) and the Personnel Armored System for Ground Troop (PASGT) helmets do not have blunt impact/TBI helmet standards, McEntire et al. (2005) subjected the ACH and PASGT mounted on a headform to blunt impact testing in accordance with Federal Motor Vehicles Safety Standards for Motorcycle helmet testing with ANSI-Z90.1 procedures and standardized head form type C (of mass 5kg) onto which the helmet configurations were fitted. The head forms were dropped from a vertical monorail drop tower and the acceleration was measured with a single axis accelerometer at the center of gravity of the head form.

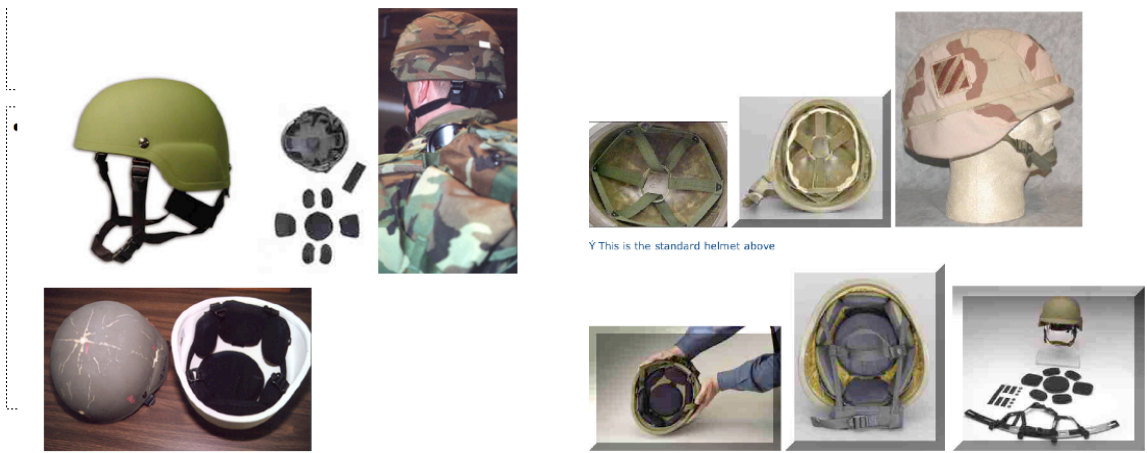

Figure 3.6: Left: MICH for Special Forces with standard issue removable cushioning pads. Right: PASGT helmet with its removable pads, which achieve approximately the same blunt impact standard as the MICH. Both are insufficient to protect against TBI to acceptable levels.

McEntire et al. 2005, determined the peak acceleration attained and the mean acceleration attained for a series of impacts for 10 fps and 14.14 fps impact velocities (corresponding to free-fall heights of 1.56 ft and 3.12 ft respectively.) Three standard helmet kits were tested: (i) Standard issue ACH which has seven internal foam fitting impact protection pads that cover the interior helmet surface, (ii) PAGST standard infantry helmet which includes only an internal head web and leather headband, (iii) PAGST paratrooper configuration which includes a nape strap, rear nape pads and a central parachutist impact liner (PIL) pad. The PIL is optional for paratroopers but was used in the testing. Neither the PIL nor the nape pads were originally designed with a specific blunt impact requirement.

The McEntire (2005) study considered 3 different temperatures, seven different impact sites, and multiple impacts. The authors defined 150G mean acceleration and a 300G peak acceleration as the criteria for acceptability at 10 feet per second (fps) and 14.14 fps impact speeds. Overall, the pad-fitted ACH outperformed the other two helmets. The ACH was the only helmet for which both the mean and the peak standards were met for all of the impact experiments at 10 fps impact speed. The PASGT with the internal liners was a close second. The PASGT without internal liners failed both mean and peak acceleration requirements. Most noteworthy is that all three helmets failed to meet the 150G mean impact standard and 300G peak standard for the more demanding but more realistic 14.14 fps impact (3.12 ft free fall).  It should also be noted that because the study considered drops with a standard head form + helmet, the standards tested are a minimum of what could be expected in the field: The tests correspond to an impact for which only the mass of the head + helmet contributes to the force which applies only for an exactly horizontal fall to the ground. Any obliquity of impact would increase the effective mass of impact.

In a follow up study, Trimble et al. 2005 identified helmet accessories to improve the blunt impact protection specifically for paratroopers and PASGT helmets. The study was the outgrowth of a 2001 USAARL solicitation to industry to submit proposed helmet impact accessories for the PASGT helmet.  Trimble et al. (2005) discuss that of 9 candidate submissions, only two performed better than the standard issue PAGST for paratroopers-- which had already failed at 14.14 fps to meet the 150G impact criterion. They then compared three systems: (1) standard issue PASGT paratrooper helmet shell with nape pads and PIL (total mass of head form + helmet kit 5kg + 1.531kg = 6.53 kg) (2) PAGST helmet shell with Skydex nape pad, PIL and a CGF Helmets (a brand name) retention harness (total mass of head form + helmet kit 5kg + 1.61kg = 6.6 kg)  (3) PAGST helmet shell with Oregon Aero pads and a CGF Helmets retention harness  (total mass of head form + helmet kit 5kg + 1.75 kg = 6.75 kg). System (2) differed from (1) in the construction and material of the nape pad. System 3 incorporated a different arrangement of helmet padding and again different materials.

The tests were performed at impact velocities of 10, 14.14, and 17.32 fps. (corresponding to free falls from 1.56 ft, 3.12 ft and 4.69 ft respectively) They tested impacts onto a curved anvil at 4 locations on the helmet. Overall, system 2 and 3 showed improvements in reducing the mean peak accelerations of impact over system 1 by 24.6% and 28.1% respectively.  These numbers are averaged over the four impact sites, all of which showed generally substantial improvements when the helmets were fitted with the newer impact inserts. Concern was raised as to whether this performance enhancement would remain in extreme temperatures or wet environments but this was not tested.

Most important however, is that the none of the new helmets could meet the 150 G peak acceleration maximum from the most realistic drop height considered: 4.69 ft with 17.32fps impact speed. This problem is exacerbated because: (1) The 150 G peak acceleration standard for the 2-4 millisecond impact time scales associated with the drop tests is already too large an acceleration to allow based on the HIC -JHTC criteria: Fig. 3.8 shows where the approximate location of the combat helmets on the acceleration vs. time  graph  in comparison  to the JHTC.  (2) The JHTC itself underestimates risk of injury, as discussed in Sec. 3.2.

As a result of the type of helmet testing just described, Skydex won a contract for 120,000 helmet pads kits for the air force and 30,000 for the army. The pads take advantage of new materials and material structures within the pads to optimize weight, volume and cushioning.  Such pads are shown in the PASGT helmet in Fig 3.6. Unfortunately, despite the thorough testing and good intentions, the helmets with the

"new" padding do not adequately protect against TBI.

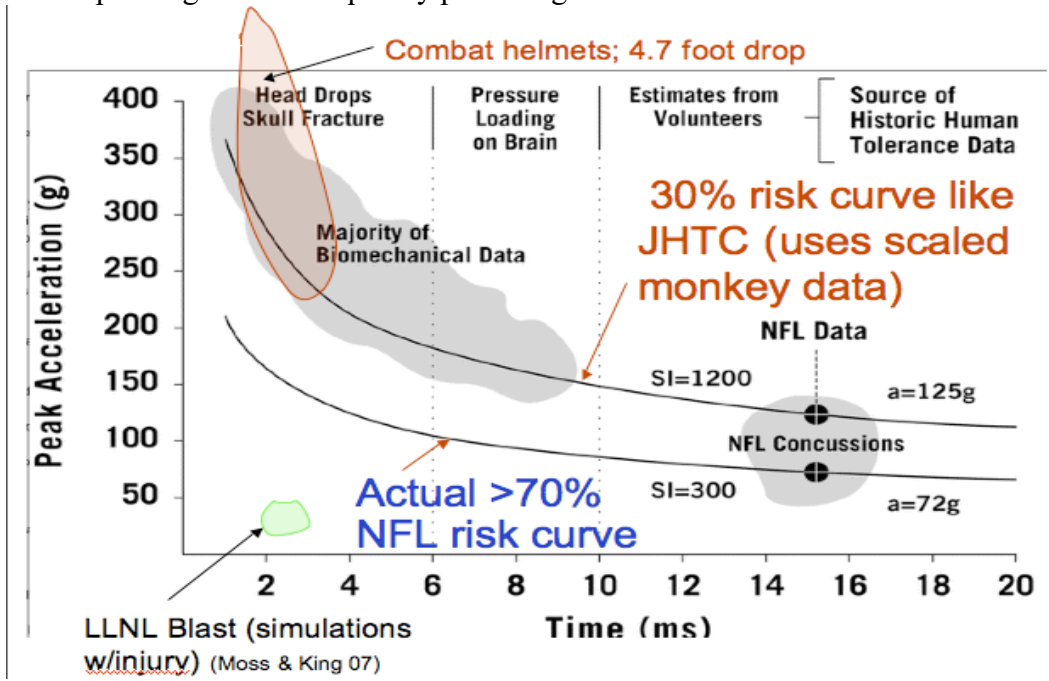

Figure 3.7: Comparison of current combat helmets, NFL football injuries, and blast injury on the acceleration vs. impact duration graph and the SI=300 and SI =1200 curves. Note that even by the JHTC standard, which is itself too weak (as exemplified clearly by NFL data) current combat helmets are ineffective.

### 3.3.3. Football: Correlation of External Accelerations with Internal Brain Stress

Perhaps the most promising approaches and accomplishments toward improved TBI protection have emerged from football helmet research. This includes technology for acceleration measurement and diagnostics, the computational modeling of brain response to forces incurred from these accelerations, and the implementation in improved helmets.

King et al. (2003) and Zhang et al. (2004) used video footage of helmet-helmet collisions in games from the National Football League (NFL) in which known concussions were incurred. The motions producing the concussions were reproduced in the laboratory using helmeted dummies, with linear and rotational accelerometers. The data measured from the reconstructed experiments were then fed as initial conditions into head impact computer simulations that include a comprehensive computer model of the human head and brain (Wayne Stead Head Model) (Zhang et al. 2001b). By analyzing the stresses on the simulated brain tissue, several proposed head injury standards in addition to the HIC could be evaluated for predictive power when compared with known injury data. From the external acceleration data measured from the videos, the external forces and energy input per unit time imparted to the head could be computed. These quantities were then correlated with the internal brain stresses from the simulations. The approach is valuable as it will ultimately allow correlations between external measures of force with internal measures of brain stress to be correlated. Input data of 22 concussions (diagnosed by team physicians) out of 53 collisions were used in these studies.

Comparing the calculated external variables, the HIC proved to be no better than the peak linear acceleration, or the head impact power (HIP) (a measure of the total kinetic energy per unit time), and little better than the rotational acceleration with respect to correlation to injury. The HIP was marginally the best correlator, followed by the peak acceleration and then the HIC, and rotational acceleration. The conceptual advantage of the HIP is that it includes the mass of the head form whereas the HIC and peak accelerations do not include the mass. However, because the impacts analyzed in the laboratory all used the same mass head forms and helmets, the role of mass in the HIP makes no distinguishable difference for these studies. A criticism is that, unlike the actual collisions for which the "effective mass" can differ for different body angles at impact, the lab experiments are set up such that the effective mass is always the same.

For the "internal" variables, the best correlators with injury emerged to be the product of the peak strain times its rate of change at the brainstem and the peak stress at the mid-brain (brain stem). (King et al. 2003; Zhang et al. 2004). The product of strain times its rate of change is actually equivalent to the change in elastic energy for a material of constant stiffness (Young's modulus) during the strain. The peak strain is the peak force per unit area. The threshold tissue strain values for brain injury are likely universal and largely independent of the external circumstance and effective mass. However, the actual stress incurred by the brain does depend on the external force and the effective mass of the impactor.

A summary of external variables and internal variables an thresholds is shown in Figure 3.7 (augmented from Newman's talk p165 in Fenner et al. 2005). We have added the peak stress in units of kPa and in atmospheres. Being that 1 atmosphere corresponds to ambient pressure and of normal ground level, air particularly noteworthy is that the injury threshold represents only 10% increase in the force per unit area.

From these analyses an "intuitive" result emerges that warrants further work: The rate of change of elastic energy of the brain tissue and the peak stress on brain tissue correlate with the rate of change of external energy imparted to the head and the peak external force imparted to the head. A systematic correlation between external and internal measures would be a powerful tool for helmet manufacturers if they need only test external variables to know that their helmets are protecting internal brain stress.

| Predictor Variable | Threshold Values for Likelihood of MTBI | | |
|---|---|---|---|
| | 25% | 50% | 75% |
| $A_{r\ max}$ (m/s$^2$) | 559 | 778 | 965 |
| $R_{r\ max}$ (rad/s$^2$) | 4384 | 5757 | 7130 |
| $HIC_{15}$ | 136 | 235 | 333 |
| $\varepsilon_{max}$ | 0.25 | 0.37 | 0.49 |
| $d\varepsilon/dt_{max}$ (s$^{-1}$) | 46 | 60 | 79 |
| $\varepsilon \bullet d\varepsilon/dt_{max}$ (s$^{-1}$) | 14 | 20 | 25 |
| Stress at midbrain (brain stem) | 6 kPa | 8 kPa | 10 kPa |
| | 0.06 atm | 0.08 atm | 0.1 atm |

. Figure 3.8: Comparison of external threshold values at the corresponding probability for concussion in NFL players from a sample of 23 impacts (King et al. 2003; Zhang et al.

2004; Newman et al. 2005). Table augmented from (Newman's talk in Fenner et al. 2005). From top to bottom, the rows are: peak acceleration (1G ~ 10m/s$^2$); peak angular acceleration; HIC for 15 ms; maximum strain; maximum strain rate; product of maximum strain and strain rate (proportional to change in elastic energy of tissue); stress at the brain stem in units of kPA and atm. Note that the maximum strains also occur at the brain stem. Note that the threshold stress for concussion corresponds in equivalent units to an effective internal overpressure of less than 0.1 atm. Not shown is the Head Impact Power threshold which for 50% injury probability as a value of 12.5 kilowatts.

### 3.3.4. Football: Real Time Data and Analysis of Head Acceleration

In addition to the progress made in simulating known football injuries by analyzing frame by frame NFL video, as described above, progress has been made in real time acceleration measurements. This is important for treatment as well as helmet design. Data collection needs to be an important part of future military helmet improvements and the football enterprise has led the way in this regard.

Duma et al. (2005) provide an example study of the Virginia Tech football team using the Simbex Head Impact Telemetry (HIT) system marketed under the name `Sideline Response System" for football teams. This wireless system provides real time data from head impacts to a receiver and laptop located on the sideline. Spring mounted accelerometers keep constant contact with the head so that it is indeed the head acceleration that is measured within the helmet. The helmet unit consists of 6 linear accelerometers, which are all needed to compute the 3 linear and 3 angular components of acceleration. (Angular accelerometers could reduce the number of required sensors.) The unit also includes 1 temperature sensor, a wireless transceiver, and memory for up to 33 impacts. Impacts are triggered at 10G and recorded over a total of 40 ms, which includes a 12ms pre-trigger and 28ms post-trigger. The current system simultaneously monitors 64 players.

The impact data from the 6 linear accelerometers were combined by an algorithm that solves the non-linear equations for the impact location, the mean linear acceleration of the center of gravity, the GSI and the HIC, along with sagittal and lateral rotational accelerations. The HIT acceleration data were validated using impact tests with a helmet-equipped Hybrid III dummy instrumented with the ``industry standard" 3-2-2-2 (9 accelerometer) method for which computation of accelerations becomes linear. The HIT matched the 3-2-2-2 accelerometer array to 4% accuracy. Verification of the calculated location of impact was performed using drop tests and comparing the observed impact site to the calculated site. Agreement was demonstrated to within 1.2 cm. With the HIT System installed, the Riddell VSR4 helmets passed all football helmet standards required at the NCAA collegiate level. As of 2007, Riddell is offering a retail version of this helmet system. Fig. 3.9 shows a version of the helmet.

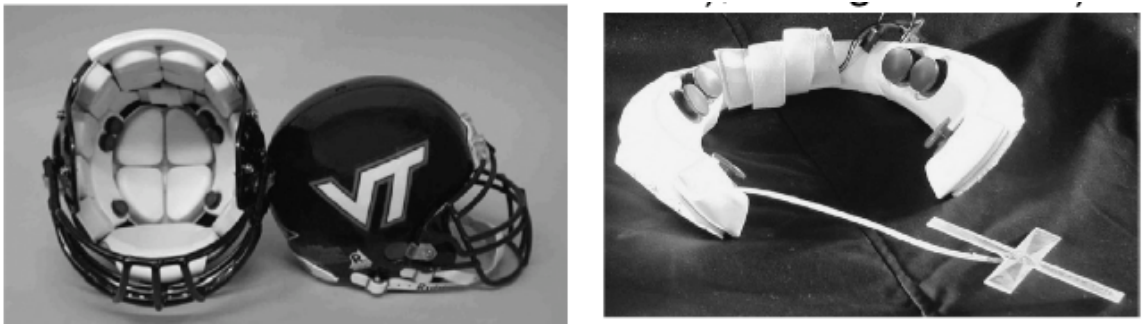

Figure 3.9: The Riddell VSR4 helmet with the 6, spring mounted accelerometers, battery pack, and antenna installed (Duma et al. 2005).

During each practice and game of 2003 season, up to 8 players of the Virginia tech Football team were monitored. To analyze the data from the HIT in combination with clinical determinations of concussions, detailed neuropsychological records were kept on each participating player so that a pre-injury cognitive baseline could be compared with post injury status. Clinical indicators monitored were headache, postural sway, confusion and retrograde and amnesia.

The HIT system recorded 3312 head impacts with 1198 occurring in 10 games. Impacts were analyzed from 38 different players covering a range of positions. For all impacts, the average peak acceleration was 32G with a range between 1G and 200G. Approximately 90% were less than 60G. Ironically, only five concussions occurred over the course of the season and only 1 occurred for a player being monitored with an HIT. This player incurred 33 impacts in a particular game and it was the second impact that produced the concussion. The peak acceleration was 81G, SI was 267, and HIC was 200. The x-axis and y-axis rotations were 5600 rad/s$^2$. The concussion was subsequently determined to be grade 1 or grade 2.

The average linear acceleration was 32 (+/- 25) G for the impacts that did not result in a concussion. Pellman et al. (2003), using videos of NFL games and reconstructing the motions using the Hybrid III dummies, found that the average linear acceleration for concussed players was 98(+/-28)G. The single Virginia Tech concussion in Duma et al. (2005) reported above is consistent within the reported error bars of this value. However, Duma et al (2005) measured 583 impacts with linear accelerations having a peak above 70G. Pellman et al. (2003) found that 50% of NFL players with linear accelerations above 78G (see Fig 3.9) seemed to incur concussions. A related inconsistency between Pellman et al. (2003) and Duman et al. (2005) is that Pellman (2003) found the tolerance minimum SI and HIC for concussions were 300 and 250 respectively, whereas Duma et al. (2005) found 71 non-concussive impacts above this supposed SI tolerance level and 55 above this supposed HIC tolerance level. Since only 1 player incurred a concussion out of 3312 total impacts, these (71+55=) 126 impacts are significant.

The statistical analyses of King et al. (2003) and & Zhang et al. (2004) also suggested that MTBI has a 75% chance of occurrence in NFL players when the HIC exceeds 333 (Fig 3.8), the linear accelerations exceed 98G, and the angular acceleration exceeds 7130 rad/s$^2$.  Only the rotational acceleration of the impact that produced the 1 concussion in Duma et al. (2005) exceeded its supposed tolerance level.

The apparent inconsistency between the threshold level for MTBI inferred by Zhang et al. (2004) and Pellman et al. (2005) when compared to Duma et al. (2005) again highlights that the key predictors for TBI have not yet been clearly understood.  It should be emphasized that the brain tissue stress thresholds of Zhang et al (2004) were not compared to the data in Duma et al. (2005). This is significant because it is the only index for which the force rather than the acceleration is quantified and independent of mass. It is possibly that one source of the differences between the NFL and collegiate player results are that NFL players have a larger mass and exert more force. For impacts of a given acceleration there is more force applied to the head for a more massive player.

### 3.3.5. Football:  Do New Helmets Reduce Concussions? Need More Independent Studies of Helmet Testing and Rigorous Improvement of Standards and Measures, and Fundamentally New Designs to Mitigate Rotational Acceleration

Despite present uncertainties in the tolerance measures and standards, studies like those described in the previous sections could in principle lead to success in improved football helmet design. However the recent claims of such are inconclusive without further study.

Viano et al. (2006) claim a 10-20% reduction in NFL concussion risk with newly designed football helmets like the Riddell Revolution compared to the older Riddell VSR-4. In principle, newer football helmets reduce concussions by using more energy-absorbing padding low on the side and back of the helmets, and around the ears. These are the primary areas where NFL players are experiencing collision inducing concussions. The size and coverage of the side of the helmet are being increased, providing more space for better liner materials such as vinyl nitrite. The benefits of the newer helmets have also been purported to apply to high school players. Collins et al. (2006) argue that the new Riddell Revolution helmet has reduced the relative probability of concussions by 31%. Athletes wearing "standard helmets" had an annual concussion rate of 7.6% compared to 5.4% for those wearing the newer helmet.  However, the standardization of identifying concussions in youth football has been questioned and the helmets to which the supposed 31% improvement applied were themselves not standardized; many youths use refurbished helmets so the comparison is not cleanly related to an improvement over a specific helmet.  Figure 3.10 which appears on the Riddell website gives an incorrect impression that the comparison is made to a specific helmet.  See Schwarz (2010) for a summary of the related important issues.

One important point is that the difference between the older and newer helmet show in Fig 13 does not represent a paradigm shift. Complementarily, a new paradigm for helmet design has emerged from Xenith LLC, culminating in their "Xenith X1" helmet. A photo is shown in Fig. 3.11.  The paradigm shift occurs in the design of the 18 plastic black shock absorbers that replace foam padding. These shock absorbers are air filled but have a small hole at their center. For low acceleration impacts, the air leaks out relatively laminarly (smoothly, and at pace with the impact). However, as the impact power increases, the force on the shock absorber increases. A strong force on the air filled shock absorbers generates turbulence inside the absorbers that inhibits the escape of air. For high power impacts therefore, the deflation is slower, tuned at the right rate to gently cushion the impact. This allows the helmet to be effective at both high and low power impacts.

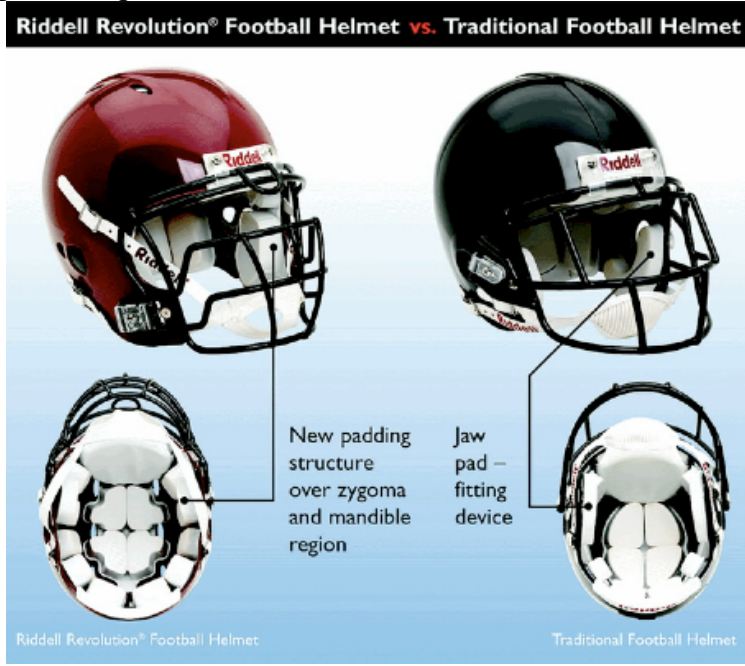

Figure 3.10: The Riddell Revolution helmet (left) compared to the early VSR4 type design. Note the increased padding, but also the fact that there is no substantial "paradigm" shift in design between the two helmets. Contrast this with figure 3.12.

The Xenith X1 helmet has received NOCSAE approval and has outperformed traditional foam padded helmets in the same drop tests to which all football helmets must be subjected. In particular, although the NOCSAE requires helmets to satisfy SI < 1200 for when the helmet is fitted to a standard head form and a 60cm drop test is performed (see Fig 3.8, and discussion of drop tests in section 3.3.2). This new helmet averaged 340 which comfortably meets the standard (which, by the way, is too low) and also performed better than traditional designs. In particular, testing has shown that the disks can withstand hundreds of impacts without any notable degradation. This is a dramatic improvement on standard designs that degrade after repeated impacts. All this bodes well for further study of this new paradigm, and potential application to the military.

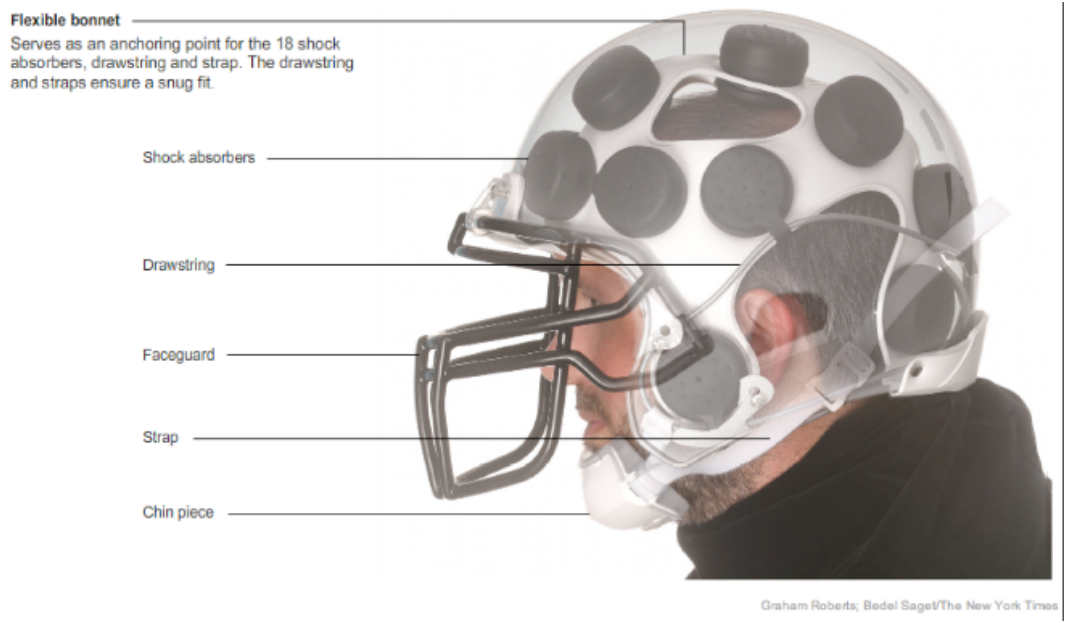

Figure 3.11: The Xenith X-1 helmet with the 18 air cushion shock absorbers described in the text (Schwarz 2007).  This is a true paradigm shift in helmet design and is likely to make a substantial improvement in TBI protection from even the Ridell Revolution in Fig 3.11 as the design is refined.

The football helmet research community has led the way with new technology, creative real time diagnostics, modeling, and helmet design in the effort to protect against impact induced TBI.  The approach has largely been empirical however, and weaknesses remain in their interpretation of the data. For example, the NOCSAE standard (HIC =1200) is too weak based on the measured threshold HIC for concussions (HIC=150) and the data analysis and modeling still do not incorporate the effective mass of the impacting player that varies depending on the angle of body impact. Although these shortcomings limit the precision with which football helmet impact data can be directly used for military helmet design, the technological approach to data collection and theoretical modeling, and the new paradigm for helmets TBI protection offer useful strategies and ideas for combat data collection and military helmet redesign.

One major conceptual flaw in all standard football helmets that has yet to be addressed as of 2011, is that they do not mitigate rotational acceleration nearly as effectively as they would if the helmet shell could move much more freely about the head with respect to inner protective cushioning.  That freedom would allow the shell to take up significant rotational energy and rotational acceleration, reducing the amount transmitted to the head and brain. This type of design should be considered an essential fundamental design change in  the next generation of helmets.

**3.4 Blast vs. Impact Injury:  Insight and Promise from New Numerical Simulations**

Our focus above on impact injury is relevant even for blast waves from explosions, which cause most of the TBI for combat soldiers in Iraq and Afghanistan.

This is because the blast presents three mechanisms for TBI usually termed as follows: (1) Primary: This is injury caused by the direct effect of overpressure (Fig 3.12) that occurs as the blast wave encounters the solider. (2) Secondary: This occurs from shrapnel propelled toward the soldiers head. (3) Tertiary: This occurs as the solider is propelled and the head hits the ground or other object—impact injury.  Despite the terminology, in the combination of (1) and (3) is likely to be even more common than the combination of (1) and (2) because the force of a blast that can cause closed TBI is enough to knock a person over such that they will hit their head.   As we have seen in Fig 3.8, the equivalent internal overpressure (force per unit area) for brain injury in the head is about 0.1 atm as inferred from the concussion threshold from numerical simulations (Zhang et al. 2004). The ambient overpressure must be larger than this overpressure to achieve such a value inside the head.  Fig. 3.12 (King and Moss, personal comm.) shows the magnitude of blast ambient overpressures that correspond to physical effects.  Already at 0.1 atm external overpressure a person is knocked down. (This is calculated simply by determining when the force per unit area on typical body, multiplied by the body cross sectional area, equals the weight). In general therefore, blast injury should be thought of at least a 2 stage process: overpressure, followed by impact.

| Blast Overpressure | Physiological Effect |
| --- | --- |
| 0.2 psi (~0.01 atm) | Minor Ear Damage |
| 1 psi (~0.1 atm) | Knock a Person Over |
| 5 psi (~0.34 atm) | Eardrum Damage |
| 15 psi (~1 atm) | Lung Damage |
| 35 psi (~2.4 atm) | Fatalities Possible |
| 65 psi (~4.4 atm) | Fatality Almost Certain |

Figure 3.12: Approximate external overpressures and their associated effect..
(King and Moss, personal communication).

The basic physics of a blast is summarized in Fig. 3.13, which represents the evolution of pressure at a point through which the blast wave passes at a fixed distance from the explosion.  The chemical energy of the explosion is converted into thermal energy of gasses ejected into the air.  These collide with air molecules and compress the air faster than the air can adjust to disperse the pressure disturbance.  The high pressure shell represents the blast front which propagates radially outward from the explosion site.  There is both an increase in the isotropic local pressure (force per unit area) measured in a frame co-moving with the blast, as well as an additional dynamical pressure associated with the coherent motion of the blast wind itself.  Initially, an observer feels a rise in the isotropic pressure and an increase in dynamical pressure directed outward. But because the blast wave has ploughed out the air from the region through which it has propagated, an under-pressure follows the initial overpressure. This explains why the top curve in Fig 3.13 rises to a peak then falls below zero. Eventually, the air refills in the cavity and the pressures fall back to the ambient level. During the refilling, the direction of airflow is reversed. The dynamical pressure is now directed backward from the initial dynamical pressure, however, when its magnitude is taken it becomes a positive quantity. This explains the arrows beneath the bottom plot in Fig. 3.13, and explains why all peaks in that plot are positive.

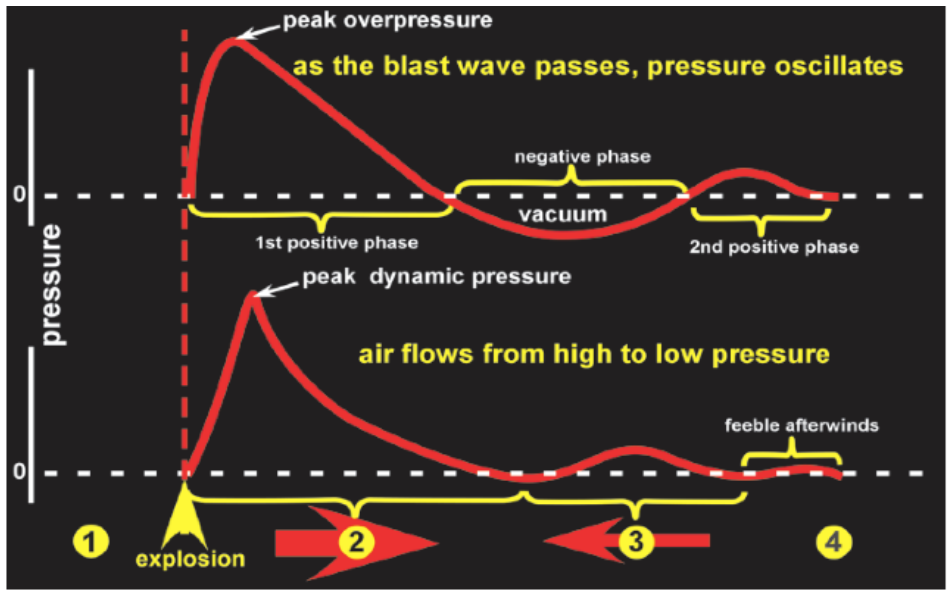

Figure 3.13: The standard pressure vs. time profile for a blast wave as measured from a fixed position that the blast passes through (Taber et al. 2006). The top curve shows the magnitude of the static overpressure relative to the undisturbed background value defined as the overpressure measured through a plane parallel to the direction of the blast motion. The bottom figure shows the magnitude of the dynamic pressure, measured as total force per unit area through a plane perpendicular to the direction of blast motion, minus the static overpressure.

As exemplified by the previous subsections, the most relevant work on helmet design for closed TBI protection has to date been the blunt impact testing from impacts. As inadequate as this testing has been and as poor as the present combat helmets are for TBI protection, even less is known about how the helmets protect against the blast overpressure injury. The relative importance and nature of overpressure induced vs. impact induced TBI from blasts is not well resolved and improved protection against both is needed.

Recently, Moss et al. (2008) have demonstrated potential progress toward understanding both overpressure and impact from effective use of high resolution numerical simulations with a state-of-the-art code ALE-3D at LLNL. The code was originally designed to support analysis of nuclear weapon explosions and the subsequent blast wave interaction with structures.

The basic set up of a typical blast and head form interaction simulation is shown in Fig. 3.14. In a canonical run, an explosion the equivalent of 5lb of C4 is detonated into standard ambient ground level atmospheric pressure and temperature conditions and the model head form is placed 15 feet from the blast. The overpressure experienced by the head form depends on the distance and the power of the blast, so the same overpressure can also be achieved by different combination of explosive charge weight and distance.

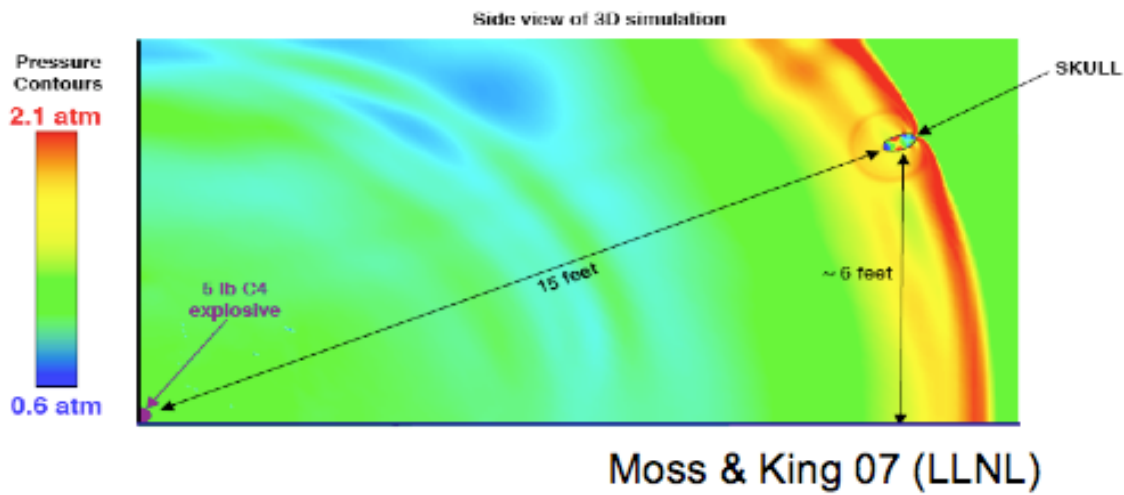

Figure 3.14: The basic numerical simulation set up (Moss and King 2007, personal communication). A head form is placed at 15 feet from the explosion and the blast wave is tracked as it propagates into the air and then passes by the helmet. High resolution allows ~50 "cells" of resolution across the helmet width, so that internal stresses and pressures inside the model brain can be measured. See figure 3.16 for an inside look at the head.

In the present preliminary studies, the head form is very simplistic, consisting of an ellipsoidal quasi-rigid shell representing the skull and uniform soft-tissue interior representing the brain. The simplicity of the head model is both a limitation and a strength: To understand the fundamental principles of how external forces produce internal brain stresses, it is useful to start with a simple physical model with only very basic features. If this can be understood, then as more complex features are included one can separate the role of the details from the principles that determine the behavior of the simplest system. Also, by comparing different helmets for the same simplified head model, it is possible to determine which helmets minimize the stresses inside the head form for a given set of external forces. For such comparative studies of helmet effectiveness, the accuracy of the internal head model is not essential. In contrast, to predict clinical measures of specific injuries that correlate with real brain scans, and to understand the biological nature of the injuries more deeply, a realistic brain model is indeed needed. Ultimately, both directions of work are possible with the simulation codes available.

The stresses and pressure gradients measured within the simulated head form can be measured and correlated with the external blast wave power. In addition, the magnitude and geometry of pressure gradients and stresses within in the brain can be compared with those resulting from a head impact using the same computer code. The general differences that emerge between overpressure injury and impact injury even for the simple head model are already beginning to suggest identifiable patterns for the location of maximum stresses and pressure gradients and the potential for correlation with clinical symptoms (Figs. 3.15 and 3.16).

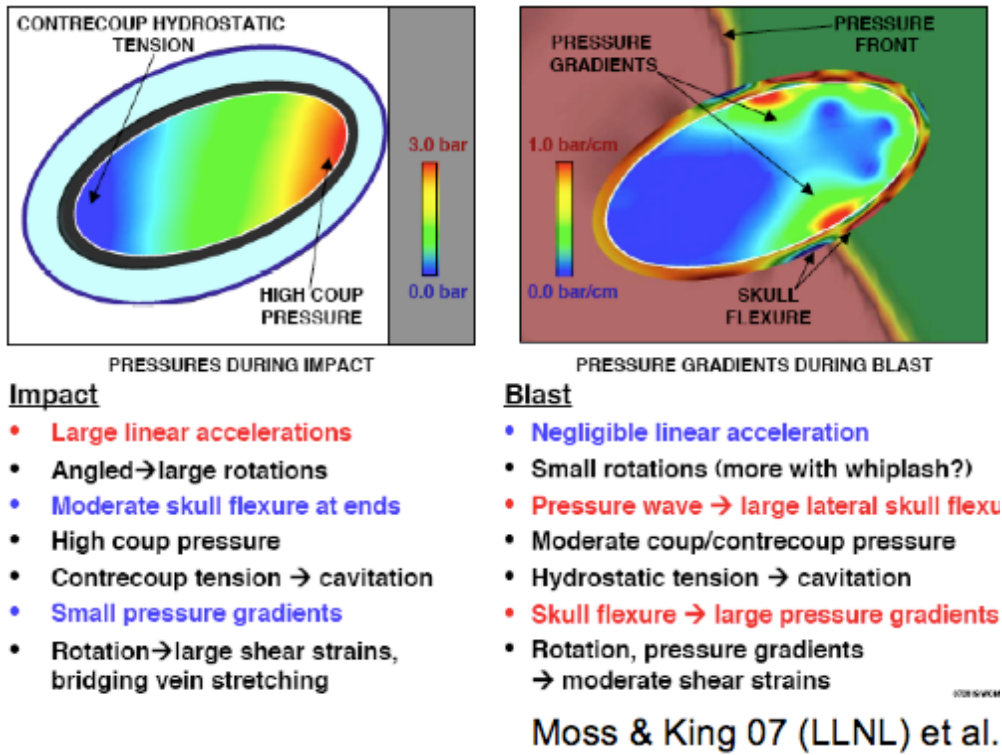

Figure 3.15: From a viewgraph of Moss & King 07; sample comparison of pressure and pressure gradients for a helmeted head incurring an impact (left), and an non-helmeted head experiencing a blast induced overpressure (right). Pressure is displayed in the left panel and pressure gradients in the right panel. For present purposes the specific numbers are not important. Rather, simply note how using the simulations can in principle distinguish between the pressure and stress distributions within the brain for impact vs. overpressure injury. More work is needed to refine and better utilize this approach.

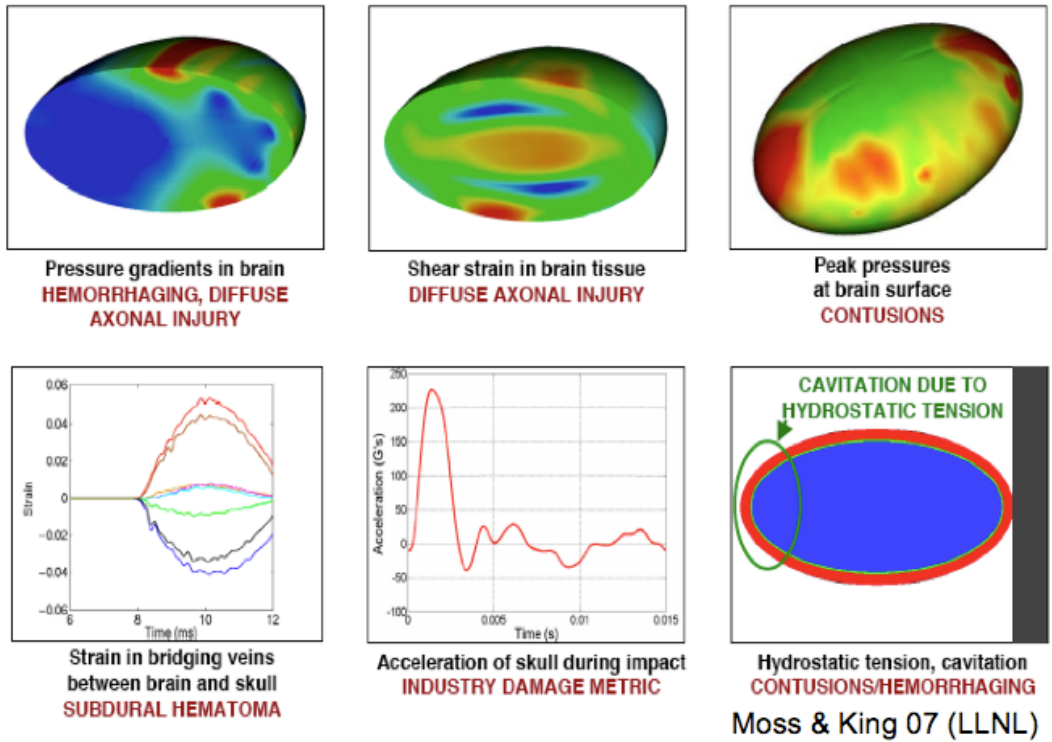

Figure 3.16: Examples of how state of the art simulations reveal important information about TBI from impact and overpressure. Top row: Diagnostics for blast overpressure injury sites in the brain that simulations can in principle predict. The red areas in the top left figure indicate the regions where the pressure gradients are maximized. In the middle figure, blue and red show regions of large oppositely signed shear strain; both are regions where stresses on brain tissue are largest. Peak pressures are shown in red in the top right figure. Bottom left: time evolution of strain on the veins between brain and skull. Bottom middle: Time evolution of acceleration from the head form impact. Bottom right: Location of potential contusion due to cavitation from impact. As a head form impacts, the brain crashes forward into the skull and deforms. CSF fills in the space from which the brain was displaced and as the brain rebounds, crashing back into the CSF fluid. That interaction is like a "belly flop" dive into a swimming pool.

Preliminary results from simulations of a blast wave propagating into a minimalist head model reveal that as the pressure wave passes by the head, it drives both pressure and shear waves into the head via skull flexure. These shear and pressure waves occur independent of the bulk motion of the head. The center of mass of the head could be kept fixed and the skull flexure induced injury can still occur. When the head is allowed to move freely, the simulations have shown that the bulk acceleration of the head in the blast is negligible (green area in Fig. 3.7) even when the internal overpressure from the skull flexure well exceeds the threshold for injury suggested by the bottom row of Fig. 3.9. This important conclusion highlights the promise of simulations even with simple head models when effective questions are asked: We have now already learned that the injury from blast overpressure has little to do with bulk acceleration and so standards targeted to reduce the HIC or SI for a given helmet from a fixed drop height are not directly addressing a quantity that determines blast overpressure injury. That being said, internal cushioning that reduces impact injury may also reduce overpressure injury by damping the waves that would otherwise propagate into the head from skull flexure. This is an important direction for future research—finding the internal cushioning that optimizes the needed protection against both overpressure and impact injury.

Having emphasized that blast waves will cause both overpressure and impact injury and that protection against both needs dramatic improvement, it is possible that the overpressure injury actually finds its way to the brain through a path other than the head. Although the head has the most surface area in contact with the brain, the skull is relatively rigid when compared to other tissues and other potentially lower resistance pathways (ears, trachea). If for example, a pressure wave compresses the spinal column and sends CSF up into the brain the pressure around the brain could increase substantially and asymmetrically. It is therefore important to emphasize that for preventing the overpressure component of blast injury, helmet protection should not be the only direction of research.

There are a number of important directions and improvements that the LLNL simulation efforts need funding to pursue. Presently, these include:

- Adding a helmet to the overpressure simulations for direct comparison with the helmeted head impact simulations. If a standing soldier is knocked to the ground, the impact will occur from an effective drop height of > 1.5 meters. A comparison of the overpressure injury and impact injury on a helmeted head should determine which type of injury dominates as a function of blast power and different helmet designs.
- Include a body form attached to the head for the impact and vary the body impact angle to extract the effective mass. Traditionally, only the head + helmet masses are used in impact studies but this is insufficient as we have emphasized in previous sections.
- For both overpressure and impact simulations, it is important to test how the rate of elastic energy change and the peak stress inside the brain correlates with the rate of external energy input and the peak external force. A realistic head model, such as the Wayne State Head Model (Zhang et al. 2001b) should be added to the LLNL simulations. This would generalize the approach of Zhang et al. (2004).
- Impact simulations should be tested on a brain model that was pre-injured from overpressure. In particular, allow a single simulation to incorporate a realistic injury scenario—blast overpressure followed by an impact for a head attached to a body form.
- The aerodynamic influence of helmet and body armor shape should be tested.
- Ultimately: correlate specific brain damage with medical symptoms and identify the external measures that can be used to effectively standardize helmets.

## 3.5 Conclusions and Recommendations: Present Combat Helmets Fail to Protect TBI but Effective Short and Long Term Strategies are Available

Because soldiers that experience a blast wave from an explosion will incur overpressure associated with the blast followed by an impact as their body is propelled or knocked over, it is important to reemphasize that both protection against overpressure and impact induced TBI need to be improved.

Much of the previous work on closed head injury from military, football, and motor vehicle helmet studies has focused on impact. But the measures for impact induced TBI are inadequate and helmet standards resulting from these measures are too lax. Current combat helmets fail to protect TBI at even the level of football helmets. A "quick fix" improvement would be to reduce the allowed peak acceleration incurred by a standard helmeted laboratory head form down to 50-150 G for a 5 foot drop and 2ms impact duration, rather than the 250-400G range that current helmets satisfy. This change could likely be accomplished by adding 50% thicker cushioning of the same material immediately, or seeking a more efficient cushioning that would keep the same volume with the extra protection. Football helmets already do a better job, and the football helmet industry leads the way for helmet design and real time data collection on head motion.

However the football standard itself is too lax and conceptually inadequate since it measures only the acceleration, not the impact energy rate of change to the head form or the force on the head which both depend on the effective mass of impact. Since force depends on mass, and current blunt impact measures such as the HIC and SI depend only on acceleration and time, theses measures are inadequate. The effective mass that determines the impact force depends on the total mass along the line perpendicular to the surface of impact at the impact site. If the body incurs a head-first fall vertically for example, the entire body+ equipment mass comprises the effective mass. Including this conceptual improvement requires a much more stringent helmet standard as laboratory testing that has determined the SI and HIC standards currently used are based on laboratory tests of ~6kg masses (head form + helmet).

The situation is even worse for overpressure induced TBI standards: There exist no standards for TBI protection from overpressure injury. This frontier warrants a substantial effort.

Among the shortest paths to dramatic improvements for both impact and overpressure protection is a focused use and improvement of the simulation studies discussed in section 3.4, with companion experimental studies where possible. We estimate that significant improvement in helmet design for TBI protection could be expected within 5 years even at $10 million per year for computational approaches to improved helmet design. This estimate is simply based on assuming that a university principal investigator can typically fund an interdisciplinary team of 3 post-docs, support staff, summer salary, and computational and overhead resources at approximately $2 million for 5 years. The $10 million would then fund the equivalent of 5 groups working toward the goal of TBI protection over 5 years. This is a small fraction of the $1 billion+ total annual medical care costs of the currently injured combat soldiers discussed in the introduction.

A list of more specific recommendations and suggestions for further consideration is summarized below:

- The Department of Defense should establish and interdisciplinary (engineering, physics, biology) research program that specifically focuses on protection of TBI via helmet design and body armor. It is important that the specific focus on protection of TBI be a separate funding program than the treatment of TBI.
- The above program should produce new paradigms for helmet inserts that maximize protection against impact AND overpressure by both damping the acceleration upon impact and damping pressure and shear waves sent thorough the helmet via helmet and skull flexure. This research should also evaluate whether a softer outer shell than currently used for helmets may be possible without reducing the effectiveness of ballistic projectile protection.
- Require immediate development of padding inserts for combat helmets that requires peak accelerations below 100 G for a 5 ft drop using the same drop tests that produced a 150G standard on the MICH helmet for a 1.5 foot drop. These inserts should be mandatory for all combat soldiers.
- Pad the inside of vehicles subject to blast exposure in combat: The cushioning required to prevent head injury from impact should be both internal and external to the helmet where possible
- Combat field data are needed. Accelerometers like the Simbex package used in the new Riddell football helmets (and used in the Viano et al. 2005 study at Virginia Tech) are needed for military application. Pressure sensors on the body and in the helmet are also needed to evaluate whether a solider incurred a blast and to measure differences in overpressure inside and outside the helmet.
- Internal and external measures and thresholds for TBI must be developed and correlated. Internal thresholds represent the magnitude of stress or pressure on the brain tissue itself that causes injury. The amount of stress delivered to the brain is determined by the external forces on the head and the rate of energy input (HIP) to the head, but the threshold of stress on brain tissue for injury is an independent quantity. Numerical simulations that combine blast overpressure and or impact forces on a realistic head model should be to specifically correlate the external and internal stresses and determine which helmets, for a given external measure, minimize the internal stress.
- In particular, a potential bivariate correlation between (1) the Head Impact Power (HIP, a measure of the energy per unit time externally imparted to the head) and the peak external acceleration with (2) the peak stress in the brain and the peak rate of change of elastic energy of brain tissue should be tested via simulations.
- Whether TBI from blast overpressure injury predominately penetrates through the skull or along a different pathway needs to be tested. Aerodynamics of the solider and body armor should be evaluated to assess whether improving such can mitigate pathways of the blast overpressure injury through the body to the skull.
- Possible innovations that include airbags in the helmets, and inertial switch tethering between the helmet and body should be investigated.

- Require neck-strengthening exercises as part of basic training to improve the coupling of the neck to body for impacts. NFL data has shown that stronger necks reduce the incidence of concussions.
- Ultimately, with a realistic head model, the peak physical forces can be correlated with specific clinical TBI symptoms. This is the ultimate long term goal.